\documentclass{mn2e}
\usepackage{times}
\input{psfig.sty}

\begin{document}

\title[Radio/X-ray timing correlations in NS and BH XRBs]
{Correlation between radio luminosity and X-ray timing frequencies in neutron
star and black hole X-ray binaries}
\author[S. Migliari, R.P. Fender \& M. van der Klis]
{S. Migliari$^1$, R.P. Fender$^{1,2}$, M. van der Klis$^{1}$
\\
$^1$ Astronomical Institute `Anton Pannekoek', University of Amsterdam, and
Center for High Energy Astrophysics, Kruislaan 403, \\ 
1098 SJ, Amsterdam, The Netherlands.\\ 
$^2$ School of Physics and Astronomy, University of Southampton
Hampshire SO17 1BJ, United Kingdom\\
}

\maketitle

\begin{abstract}

We report on correlations between radio luminosity and X-ray timing features
in X-ray binary systems containing low magnetic field neutron stars and black
holes. The sample of neutron star systems consists of 4U~1728-34, 4U~1820-34,
Ser X-1, MXB~1730-335, GX~13+1, the millisecond X-ray pulsars SAX~J1808.4-3658
and IGR~J00291+5934, and these are compared with the black hole system GX
339-4. The analysis has been done using data from pointed observations of the
Rossi X-ray Timing Explorer coordinated with radio observations. In the
neutron star systems the radio luminosity $L_{R}$ is correlated with the
characteristic frequency of the $L_{h}$ Lorentzian component detected
contemporaneously in the power spectrum, and anticorrelated with its
strength. Similarly, in the black hole system GX 339-4 $L_{R}$ is correlated
with the frequency of the $L_{\ell}$ component in the power spectrum, and
anti-correlated with its strength. The index of a power-law fit to the
correlation is similar in both cases, $L_{R} \propto
\nu^{\sim 1.4}$ and $L_{R} \propto (rms)^{-2.3}$. At lower timing frequencies,
the radio luminosity is further found to be correlated with the characteristic
(break) frequency of the $L_{b}$ component of the power spectra in the neutron
stars and, marginally, with the equivalent break frequency in GX 339-4. We
briefly discuss the coupling between the innermost regions of the accretion
disc and the production of the jet and, from the behaviour of the ms accreting
X-ray pulsars, the possible role of the NS magnetic field.

\end{abstract}


\section{Introduction}
\begin{figure*}
\begin{tabular}{l}
\psfig{figure=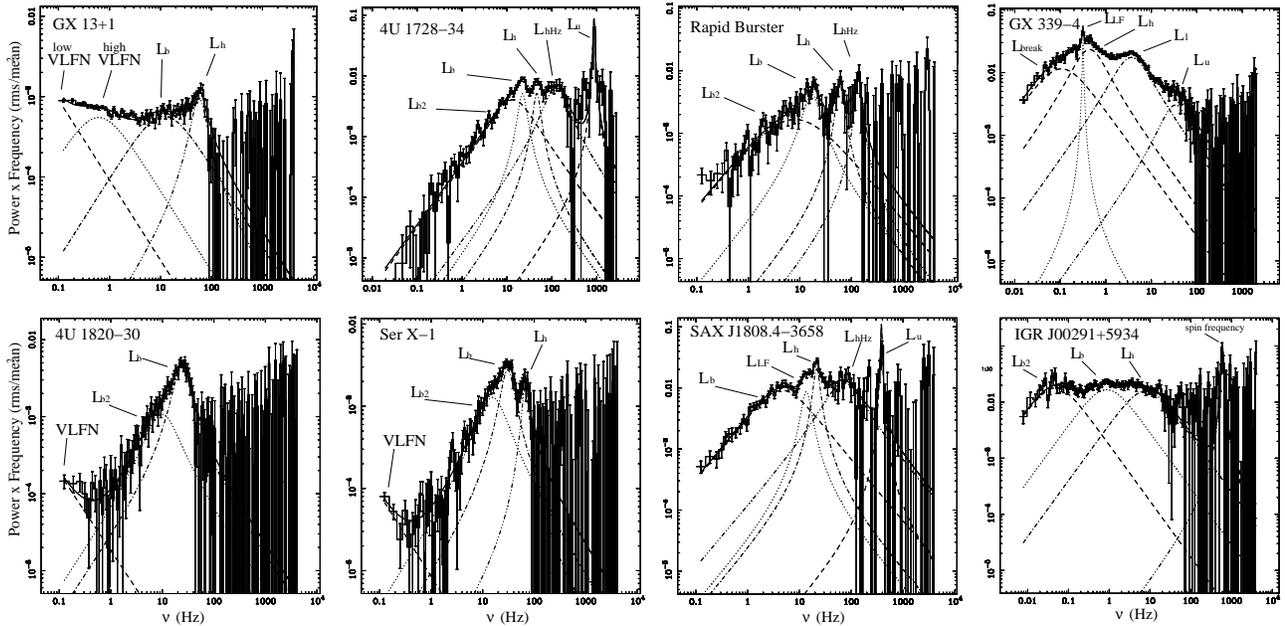,width=17cm,angle=0}\\
\end{tabular}
\caption{Power density spectra with the fitting multi-Lorentzian model
of the NS X-ray binaries analysed: GX~13+1, 4U~1728--34, MXB 1730--335 (the
Rapid Burster), 4U~1820--30 and Ser~X-1, SAX J1808.4--3658 and
IGR~J00291+5934, and a typical power spectra of the BH GX~339-4 during hard
state.}
\end{figure*}
Jet production in X-ray binaries (XRBs) is strongly related to the properties
of the accretion disc.  Recent works have established a link between the disc
X-ray power and the jet radio power in black hole (BH) and neutron star (NS)
XRB systems. In BH XRBs a non-linear correlation links the radio and the X-ray
luminosity when in the hard state (e.g. Corbel et al. 2003; Gallo, Fender \&
Pooley 2003). Accretion disc and jet theories can translate this coupling in
to a relation between the power in the jet and mass accretion rate
($\dot{M}$). Heinz \& Sunyaev (2003) have derived a non-linear relation
between the radio power of the jet observed at a given frequency, the mass of
the compact object and the mass accretion rate.
In the case of BHs, it is common use to convert the (bolometric) X-ray
luminosities in Eddington units, and infer the mass accretion rate, that is
the luminosity is a direct estimator of $\dot{M}$, although, as predicted
already by e.g. the ADAF model (e.g.  Narayan \& Yi 1994, 1995; Narayan,
Garcia \& McClintock 1997), observationally the relation between luminosity
and other $\dot{M}$ indicators is not straightforward [Homan et al. 2001; see
also Homan \& Belloni (2005), Remillard (2005), McClintock \& Remillard
(2005)]. In the case of NSs, the relation between the X-ray luminosity and
other mass accretion rate indicators is definitely not one-to-one. Parallel
tracks are observed in NS XRBs, both for individual sources and across
sources, between the centroid frequency of the kHz quasi-periodic oscillations
(QPOs) in the X-ray power spectra and the X-ray luminosity (e.g. M\'endez et
al. 1999; Ford et al. 2000; van der Klis 2001). QPOs are thought to be related
to accretion disc properties, and in particular kHz QPO frequencies are
generally interpreted as being related to the motion of matter in the
accretion disc at a preferential radius (see e.g. van der Klis 2004 for a
review). This suggests that the kHz QPOs may be a more direct
indicator, rather than luminosity, at least of accretion geometry (see van der
Klis 2001). Many works showed that the characteristic frequency of the kHz
QPOs are strongly related to the characteristic frequencies of the other
timing features in the power spectra, so that lower-frequency QPOs (much
easier to detect than the kHz ones) can be used as a proxy (e.g. Psaltis,
Belloni \& van der Klis 1999; Belloni, Psaltis
\& van der Klis 2002; van Straaten et al. 2002; van Straaten, van der Klis \&
M\'endez 2003; van Straaten, van der Klis \& Wijnands 2005).

A positive correlation between radio and X-ray flux, reminiscent of that found
in BH XRBs, has been found in the atoll-type NS 4U~1728--34 when in its hard
X-ray state [Migliari et al. 2003; see Hasinger \& van der Klis (1989) for a
classification of NSs].  Eight atolls and low-luminosity NS XRBs (compared to
Z-type NSs) have been detected in radio so far, during coordinated radio and
X-ray observations: 4U~1728--34 when steadily in its hard state (Migliari et
al. 2003), 4U~1820-34 and Ser X-1 when steadily in their soft states (Migliari
et al. 2004), Aql~X-1 during an X-ray outburst (Rupen et al. 2004), MXB
1730--335 (the Rapid Burster) during X-ray outbursts (Rutledge et
al. 1997,1998; Moore et al. 2000), the two millisec accreting X-ray pulsars
SAX~J1808.4--3658 and IGR~J00291+5934 during X-ray outbursts (Gaensler,
Stappers \& Getts 1999; Rupen et al. 2002; Pooley 2004).  The `peculiar' atoll
source GX~13+1, difficult to classify through X-ray analysis, having some
X-ray properties of Z-type and some of atoll-type sources (but more likely
identified as a bright atoll source: Schnerr et al. 2003), has also been
detected at radio wavelengths, and shows a radio luminosity similar to
(i.e. as strong as) the one of Z sources (Homan et al. 2004).

\begin{table}
\centering
\caption{Name of the source, date of the beginning of the observation and Obs. ID of the PCA/RXTE observation} 
\label{qpo-radio_tab1}
\vspace{0.2cm}

\begin{tabular}{l l l}
\hline
Source & Date & Obs ID\\
\hline
\multicolumn{3}{c}{NEUTRON STARS} \\
\hline
 4U~1728--34 &2000 May 05 & 50023-01-14-00\\
&2000 May 13 & 50023-01-16-00\\
&2000 May 21&50023-01-19-00\\
&2001 May 29&60029-02-02-00\\
&2001 Jun 01&60029-02-03-00\\
&2001 Jun 03&60029-02-04-00\\
&2001 Jun 05&60029-02-05-00\\
&2001 Jun 07&60029-02-06-00\\
&2001 Jun 09&60029-02-07-00\\
\hline
4U~1820--30 &2002 Jul 25 &70030-03-01-01\\
\hline
Ser X-1 &2002 May 27 &70027-04-01-00\\
\hline
 MXB 1730--335 &1996 Nov 6&20093-01-01-00\\
\hline
SAX J1808.4--3658 &1998 Apr 27 &30411-01-10-01\\
&2002 Oct 16&70080-01-01-04\\
&2002 Oct 18&70080-01-02-01\\
\hline
IGR~J00291+5934 &2004 Dec 6&90425-01-01-02\\
\hline
GX 13+1 &1999 Aug 1&40022-01-01-01\\
&1999 Aug 4&40022-01-02-000\\
\hline 
\hline
\multicolumn{3}{c}{BLACK HOLE} \\
\hline
GX 339-4  &1997 Feb 4&20181-01-01-00\\ 
 & 1997 Feb 11& 20181-01-02-00\\
 & 1997 Feb 18&20181-01-03-00\\
 & 1999 Mar 3& 40108-01-04-000\\
 & 1999 Apr 2& 40108-02-01-00 \\
 & 1999 Apr 22& 40108-02-02-00\\
 & 1999 May 14& 40108-02-03-00 \\
\end{tabular}
\end{table}

\section{Observations and data analysis}

We have inspected all the available X-ray observations with the Proportional
Counter Array (PCA) on-board the Rossi X-ray Timing Explorer (RXTE),
coordinated with radio observations (see Table~\ref{qpo-radio_tab1}) of
atoll-type and ms accreting X-ray pulsars with a detected radio counterpart,
i.e. 4U~1728--34, 4U~1820-34, Ser X-1, MXB 1730--335, GX~13+1,
SAX~J1808.4--3658 and IGR~J00291+5934, and of the BH GX~339-4 in the hard
state from Corbel et al. (2001). Throughout the paper we will use a `broader'
definition of atoll sources as (non pulsating) low-magnetic fields NS XRBs
accreting at low mass accretion rates (compared to Z-type NSs), which in this
specific case includes also MXB~1730--335 (the Rapid Burster). Therefore,
hereafter we will devide the sample in atolls and ms accreting X-ray pulsars.
This is a more simplified definition (see van der Klis 2005 for a more
detailed classification), but it is appropriate for the discussion in this
paper. All the observations in our sample, with dates, radio flux densities at
8.5~GHz and the best-fit values of the characteristic frequencies of X-ray
low-frequency features are shown in Table~\ref{qpo-radio_tab2}.

\begin{table*}
\centering
\caption{Name of the source, date of the beginning of the radio observations
coordinated with X-rays, flux density at 8.46~GHz, characteristic frequency of
L$_{b2}$, L$_{b}$ and L$_{h}$ (L$_{break}$, L$_{h}$ and L$_{\ell}$ for the
black hole), the distance to the source and the references. Errors on the
radio flux is $1\sigma$, errors on frequencies and rms are 90\% statistical
errors.}
\label{qpo-radio_tab2}
\vspace{0.2cm}

\begin{tabular}{l l l l l l l l l}
\hline
\multicolumn{8}{c}{NEUTRON STARS} \\
\hline
Source & Obs Date & F$_{8.5}$~(mJy) & $\nu_{b2}$ &$\nu_{b}$ & $\nu_{h}$
& D (kpc)& Ref.\\
\hline  
\hline
4U~1728--34 &2000 May 05 & $0.6\pm0.2$ & $16.14\pm3.82$ & $22.35\pm0.72$  &
$48.12\pm2.07$ & 4.6& M03,G03\\
&2000 May 13& $0.33\pm0.08$ & $5.70\pm0.66$  &$15.77\pm0.57$  & $27.84\pm1.45$ &&&\\
&2000 May 21& $0.62\pm0.10$  & $7.16\pm0.49$  &$18.43\pm0.47$  & $30.14\pm1.68$ &&&\\
&2001 May 29& $0.11\pm0.02$ & $-$            &$2.67\pm0.23$   & $15.64\pm0.53$ &&&\\
&2001 Jun 01& $0.09\pm0.02$ & $-$            &$2.36\pm0.58$   & $13.13\pm1.45$ &&&\\
&2001 Jun 03& $0.11\pm0.02$ & $-$            &$1.50\pm0.23$   & $13.89\pm1.19$ &&&\\
&2001 Jun 05& $0.15\pm0.02$ & $-$            &$1.62\pm0.18$   & $10.95\pm0.55$ &&&\\
&2001 Jun 07& $0.16\pm0.02$ & $-$            &$1.94\pm0.28$   & $11.78\pm0.71$ &&&\\
&2001 Jun 09& $0.09\pm0.02$ & $-$            &$1.37\pm0.11$   & $8.37\pm0.36$  &&&\\
\hline
4U~1820--30 &2002 Jul 25 &$0.138\pm0.035$ &$9.15\pm2.38$ &$23.42\pm0.75$ & $-$
&7.6&M04,H00\\
\hline
Ser X-1 &2002 May 27 & $0.076\pm0.015$ & $15.50\pm1.83$ &$30.56\pm0.76$ &
$67.07\pm2.11$&12.7&M04,JN04\\
\hline
MXB 1730--335 &1996 Nov 6&$0.370\pm0.045$&$8.08\pm4.04$ &$17.21\pm1.13$ &
$60.08\pm2.71$&8.8&M00, K03\\
\hline
SAX J1808.4--3658 &1998 Apr 27 & $0.8\pm0.18$&$-$ &$0.814\pm0.100$ &
$-^{a}$ &2.5&G99,Z01\\
& 2002 Oct. 16& $0.44\pm0.06$&$4.50\pm1.72$ & $14.62\pm0.44$ &$76.30\pm2.03$  & &R02\\
& 2002 Oct. 18& $0.30\pm0.08$& $-$ &$4.53\pm0.55$ &$21.43\pm0.40$  & &\\
\hline
IGR~J00291+5934 &2004 Dec 6$^{b}$ & $0.180\pm0.035^{c}$ & $0.052\pm0.006$ &
$0.69\pm0.08$ &$4.86\pm0.94$ &4&G05,F04\\ 
\hline
GX 13+1 &1999 Aug 1 & $0.25\pm0.05$ & $-$ & $4.45\pm0.66$ &$-$ &7&GS86,H04\\ 
        &1999 Aug 4 & $0.9-5.2^{d}$ & $-$ & $11.86\pm1.68$ &$56.86\pm2.63$ & & \\ 
\hline
\hline
\multicolumn{8}{c}{BLACK HOLE} \\
\hline
Source & Obs Date & F$_{8.5}$~(mJy) & $\nu_{break}$ &$\nu_{h}$ & $\nu_{\ell}$
& D (kpc)& Ref.\\
\hline  
\hline
GX 339-4  &1997 Feb 4 & $9.10\pm0.10$ & $0.102\pm0.013$ & $0.50\pm0.02$
&$3.89\pm0.09$ &$>7$&C00,H04\\ 
 & 1997 Feb 11& $8.20\pm0.10$ & $0.107\pm0.014$ & $0.43\pm0.02$
&$3.67\pm0.07$ &$$&Z04\\
 & 1997 Feb 18& $8.70\pm0.20$ & $0.085\pm0.010$ & $0.42\pm0.01$ &$3.63\pm0.06$ &$ $&\\
 & 1999 Mar 3& $5.74\pm0.06$ & $0.095\pm0.009$ & $0.60\pm0.01$ &$4.32\pm0.13$ &$ $&\\
 & 1999 Apr 2& $5.10\pm0.07$ & $0.080\pm0.015$ & $0.35\pm0.02$ &$3.11\pm0.14$ &$ $&\\
 & 1999 Apr 22& $3.11\pm0.04$ & $0.037\pm0.011$ & $0.21\pm0.06$ &$2.57\pm0.07$ &$ $&\\
 & 1999 May 14& $1.44\pm0.04$ & $-$ & $0.05\pm0.01^{e}$ &$1.41\pm0.12$ &$ $&\\
\end{tabular}
\flushleft

{\bf a:} The power spectrum of this observation allowed to identify L$_{b}$ at
$\sim0.8$~Hz, but due to the bad statistics above 4~Hz only a broad feature
can be fitted at $\sim17~Hz$ (see van Straaten, van der Klis \& Wijnands
2004).  {\bf b:} The date refers to the start of the radio observation, the
public RXTE observation we have analysed started on December 7, a few hours
after the end of the radio observations. {\bf c:} The flux density of
$0.250\pm0.035$~mJy at 5~GHz (Fender et al. 2004) has been converted to a flux
density at 8.5~GHz assuming an optically thin spectrum with spectral index
$\alpha=-0.6$ (S$_{\nu}\propto\nu^{\alpha}$, where S$_{\nu}$ is the radio flux
density observed at a frequency $\nu$). {\bf d:} The radio emission during the
flare has been converted to 8.5~GHz assuming an optically thin radio spectrum
with $\alpha=-0.6$ (e.g. Cooke \& Ponman 1991; Penninx et al. 1993; Hjellming
et al. 1990a,b; Fomalont, Geldzahler \& Bradshaw 2001). {\bf e:} the
identification is ambiguous between L$_{h}$ (more likely) and
L$_{break}$. {\bf Ref.:} M03=Migliari et al. 2003; G03=Galloway et al. 2003;
H00=Heasley et al. 2000; M04=Migliari et al. 2004; JN04=Jonker \& Nelemans
2004; M00=Moore et al. 2000; K03=Kuulkers et al. 2003 and references therein;
G99=Gaensler, Stappers \& Getts 1999; Z01=in 't Zand et al. 2001; R02=Rupen et
al. 2002; G05=Galloway et al. 2005; F04=Fender et al. 2004; C01=Corbel et
al. 2000; H04=hynes et al. 2004; Z04=Zdziarsky et al. 2004.
\end{table*}

\subsection{X-ray timing analysis}

For NSs, we have used {\it event} data with a time resolution of 125 $\mu$s
(E\_125us\_64M\_0\_1s) for the production of the power spectra of 4U~1728--34,
4U~1820--30, Ser~X-1 and IGR J00291+5934, and of 16 $\mu$s
(E\_16us\_64M\_0\_1s) for MXB~1730--335 (the Rapid Burster) and
SAX~J1808.4--3658. We used time bins such that the Nyquist frequency is
4096~Hz. For each observation we created power spectra from segments of 16s
(of 256s for IGR J00291+5934) length using Fast Fourier Transform techniques
(van der Klis 1989 and references therein) and we removed data drop-outs and
X-ray bursts from the data, but no background subtraction was performed. No
deadtime corrections were done before creating the power spectra. We averaged
the Leahy-normalised power spectra (Leahy et al. 1983) and subtracted the
predicted Poisson noise spectrum applying the method of Zhang et al. (1995),
shifted in power to match the spectrum between 3000 and 4000~Hz. We converted
the normalisation of the power spectra to squared fractional rms (e.g. van der
Klis 1995).

For the BH GX~339-4, we have used the same procedures as for NS. We used
combined {\em binned} (B\_250us\_2A\_0\_13\_Q), {\em single-bit}
(SB\_250us\_14\_17\_2s) and {\em event} (E\_250us\_128M\_18\_8s) data with a
time resolution of 250$~\mu$s for the production of power spectra and a
Nyquist frequency of 2048~Hz. We created power spectra from segments of 128s,
averaged them and subtracted the predicted Poisson noise spectrum shifted in
power to match the spectrum between 1000 and 2048~Hz.

We have fitted all the power spectra with a multi-Lorentzian model
(e.g. Belloni et al. 2002 and references therein), and plotted in the $\nu
P_{\nu}$ representation (with $P_{\nu}$ the normalized power and $\nu$ the
frequency). In this representation the Lorentzian component attains its
maximum at the characteristic frequency $\nu_{max}$. The frequencies quoted
here are therefore all $\nu_{max}$ values, where
$\nu_{max}=(\nu_{0}^{2}+\Delta^{2})^{1/2}$ ($\nu_{0}$ is the centroid
frequency of the Lorentzian and $\Delta$ the half width at half maximum),
which are comparable to the central frequencies values $\nu_{0}$ for sharp
(quality factor Q=$\frac{\nu_{0}}{FWHM}~\geq2$, with FWHM the Full Width at
Half Maximum) features (Belloni et al. 2002).

The features in the power spectra have been identified and labeled based on
the works of Psaltis et al. (1999), Belloni et al. (2002), van Straaten et
al. (2002, 2003, 2005), Altamirano et al. (2005) submitted, Linares et
al. (2005) submitted, Klein-Wolt, van Straaten \& van der Klis (2005), in
prep. (see Klein-Wolt 2004).

\subsection{The sample}
\label{sample}
{\em 4U~1728--34:} Migliari et al. (2003) reported a correlation between radio
flux densities and characteristic frequency of the X-ray low-frequency
QPOs. We used the values of the radio flux density at 8.5~GHz (taken with the
Very Large Array: VLA) listed in their Table~1. The power spectra were fitted
using one broad Lorentzian to represent the low frequency noise and the break
frequency (L$_{b}$ or L$_{b2}$), one or two narrower Lorentzians (L$_{b}$ and
L$_{h}$) below $\sim50$~Hz, a broad Lorentzian around 100~Hz (L$_{hHz}$) and
narrow Lorentzians to fit the kHz QPOs. A typical power spectrum of
4U~1728--34 is shown in Fig.~1. The identification of the features has been
made using van Straaten et al. (2002).

{\em 4U~1820--30:} Migliari et al. (2004) reported a radio detection (with the
VLA) at 8.5~GHz of the source when steadily in its soft (lower banana: LB)
X-ray state. Only on 2002 July 25, among the seven simultaneous radio/X-ray
observations of 4U~1820--30 taken every 2-3 days, the radio source was
significantly detected ($\sim4\sigma$, although an average over the seven days
of radio observations gives a $5\sigma$ level detection). The power spectrum
of the PCA/RXTE observation on this date is shown in Fig.~1. We needed three
Lorentzians to fit the power spectrum: a broad Lorentzian to fit the very low
frequency noise (VLFN), L$_{b2}$ at about 9~Hz and L$_{b}$ around 23~Hz. The
identification of the features has been made using van Straaten et al. (2002,
2003, 2005) and Altamirano et al. (2005), submitted.

{\em Ser X-1:} Migliari et al. (2004) reported the discovery (with the VLA) of
the radio counterpart of Ser X-1 at 8.5~GHz, while the source was steadily in
its soft (LB) X-ray state.  The power spectrum of the PCA/RXTE observation
simultaneous with radio is shown in Fig.~1.  We have fitted the power spectrum
of Ser X-1 with four Lorentzians: a broad Lorentzian for the VLFN, L$_{b2}$ at
$\sim15$~Hz, L$_{b}\sim30$~Hz and L$_{h}$ at $\sim 70$~Hz. The identification
of the features has been made using van Straaten et al. (2002, 2003, 2005) and
Altamirano et al. (2005), submitted.

\begin{figure*}
\begin{tabular}{c}
\psfig{figure=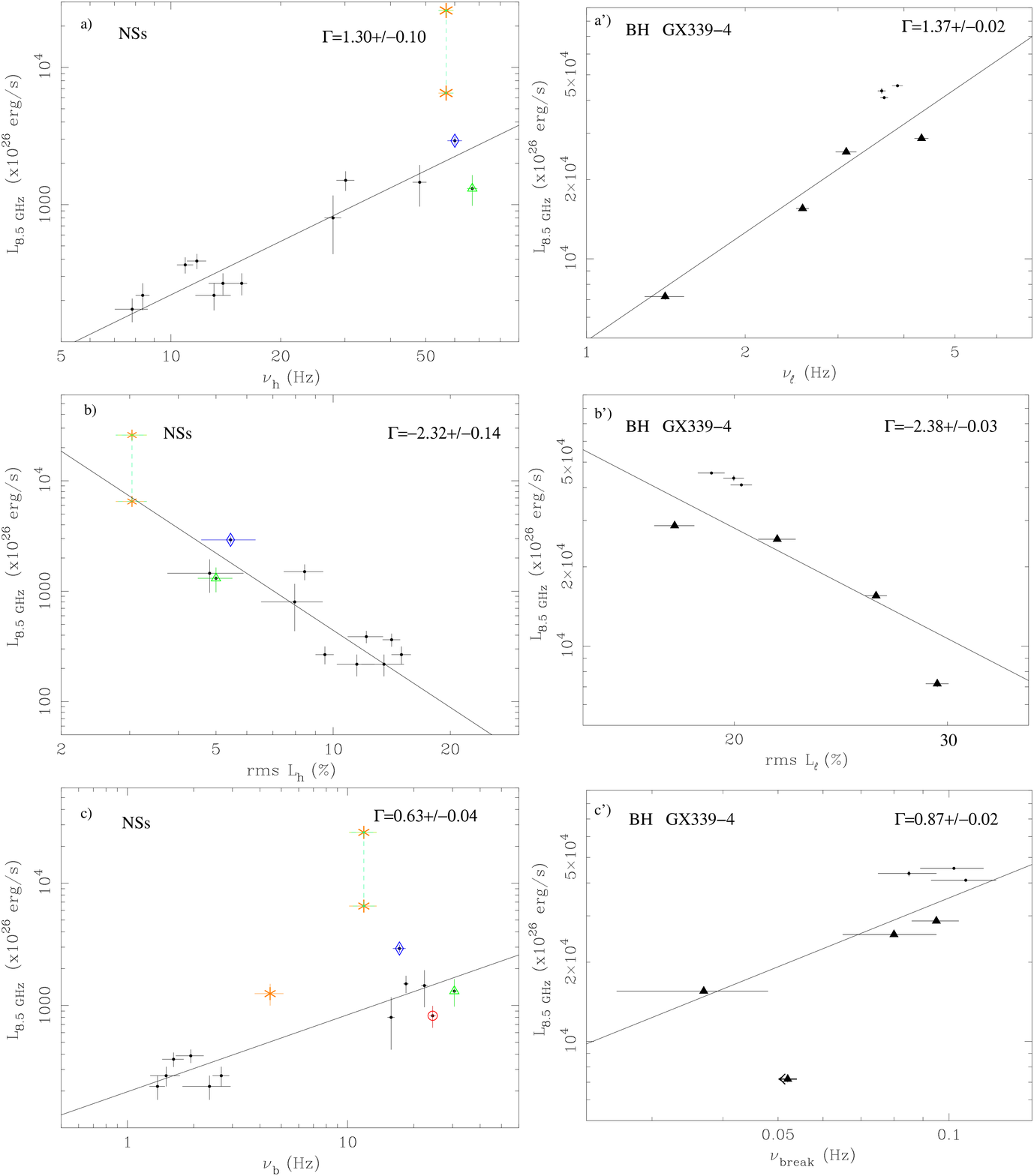,width=15cm,angle=0}\\
\end{tabular}
\caption{
{\em Top panels:} Radio luminosity at 8.5~GHz as a function of the
characteristic frequency $\nu_{h}$ for NSs (a) and of $\nu_{\ell}$ for the BH
GX~339-4 (a'); {\em Middle panels:} Radio luminosity at 8.5~GHz as a function
of the rms of L$_{h}$ for NSs (b) and the rms of L$_{\ell}$ for the BH
GX~339-4 (b'); {\em Lower panels:} Radio luminosity at 8.5~GHz as a function
of $\nu_{b}$ for the NSs (c) and $\nu_{break}$ for the BH GX~339-4 (c'). NSs
are 4U~1728-30 (dots), Ser~X-1 (open triangle), the Rapid Burster (open
diamond), 4U~1820--30 (open circle), and the `peculiar' atoll GX~13+1
(asterisks; the dashed vertical line indicates the range in radio luminosity
observed during the flare, while the timing frequency has been estimated
averaging the whole X-ray observation; see Homan et al. 2004).  In the BH
GX~339-4 we marked with dots the observations in hard state before the 1998
X-ray outburst, and with triangles the observations in hard state after the
outburst. The upper limit on the $\nu_{break}$ frequency in panel c' is the
characteristic frequency $\nu_{h}$ (see Table~1). The slope $\Gamma$ of the
fitting power-laws (solid lines; GX~13+1 has been excluded from the fit, see
\S~\ref{results}) are indicated on the top right corner of each panel (see
\S~3).}
\end{figure*}

{\em MXB 1730--335 (Rapid Burster):} Moore et al. (2000) reported the
discovery of the radio (transient) counterpart of the Rapid Burster, detected
(with the VLA) during two X-ray outbursts in November 1996 and June 1997. Due
to the large field of view of the PCA, during on-axis pointings of the Rapid
Burster the estimated flux is contaminated by the presence of the positionally
near-by bright 4U~1728--34. We have therefore analysed slew data of the
observation of 1996 November 6, when 4U~1728--34 was not anymore in the field
of view of the PCA. (The slew observations contemporaneous to the other radio
detections have not enough statistics to identify the components in the power
spectra.)  We have fitted the power spectrum with four Lorentzians: a broad
L$_{b2}$ at $\sim8$~Hz, and three narrower features at higher frequencies:
L$_{b}$ at $\sim17$~Hz, L$_{h}$ at $\sim60$~Hz and L$_{hHz}$ at $\sim135$~Hz
(see Fig.~1). The identification of the features has been made by us, using
van Straaten et al. (2002, 2003, 2005) and Altamirano et al. (2005),
submitted, based on the similarities of the power spectra to that of other
atoll sources.

{\em SAX~J1808.4--3658:} Gaensler, Stappers \& Getts (1999) reported the
discovery of a radio transient emission (with the Australian Telescope Compact
Array: ATCA)from the millisec accreting X-ray pulsar SAX~J1808.4--3658 during
the decay of an X-ray outburst on 1998 April 27. Only an upper limit on the
radio emission (lower than the detection on April 27) have been found at
8.5~GHz on April 26, suggesting a re-flaring. The power spectrum of the
quasi-simultaneous PCA/RXTE observation of April 27 allowed to identify
L$_{b}$ at $\sim0.8$~Hz, but due to the bad statistics above 4~Hz, only a
broad not reliably-identifiable feature can be fitted around $17$~Hz (see also
van Straaten, van der Klis \& Wijnands 2005). The PCA/RXTE observation on
April 26 allows to identify both L$_{b}$ and L$_{h}$ in the power
spectrum. Rupen et al. (2002) reported two radio detections at 8.5~GHz with
the VLA, during the peak of the X-ray outburst on 2002 October 16 and October
18. In the power spectra of the simultaneous PCA/RXTE observations on October
16 and 18 we identified five and six Lorentzian components, respectively:
L$_{b2}$ (only on October 18), L$_{b}$, L$_{LF}$, L$_{h}$, L$_{hHz}$ and
L$_{u}$.  The identification of the features has been made using van Straaten
et al. (2005) and Linares et al. (2005), submitted.

{\em IGR~J00291+5934:} this millisec accreting X-ray pulsar (Eckert et
al. 2004; see Markwardt, Swank \& Strohmayer 2004; Galloway et al. 2005) has
been detected in radio, with the Ryle Telescope, at 15~GHz with a flux density
of 1.1~mJy almost at the peak of the X-ray outburst (Pooley 2004). The radio
emission faded rapidly in the following days (Fender et al. 2004), suggesting
an emission from a discrete relativistic outflow, launched likely at the peak
of the X-ray outburst. We have analysed the power spectrum of the PCA/RXTE
observation of IGR~J00291+5934 on 2004 December 7, the first public
observation available with quasi-simultaneous radio observations. We could fit
the power spectrum with three Lorentzians: L$_{b2}$ at $\sim0.05$~Hz, L$_{b}$
at $\sim0.7$ and L$_{h}$ at $\sim5$~Hz (see Fig.~1).  The identification of
the features has been made using van Straaten et al. (2005) and Linares et al.
(2005), in prep.

{\em GX~13+1:} the `peculiar' and persistently bright atoll source GX~13+1 has
been observed simultaneously in X-rays (with the PCA/RXTE) and radio at 5~GHz
(with the VLA) by Homan et al. (2004). They reported a relation between the
position on the X-ray colour-colour diagram (CD) and the radio emission
behaviour (i.e. flaring emission in the harder state). The classification of
GX~13+1 as atoll- or Z-type NS is controversial. GX~13+1 shows many X-ray
properties common to both classes, although more similar to atoll sources
(Schnerr et al. 2003), but radio properties reminiscent of Z sources (Homan et
al. 2004). We have converted the radio flux densities from 5~GHz to 8.5~GHz
assuming a flat spectrum during the steady emission in the soft state, and
$\alpha=-0.6$ (S$_{\nu}\propto\nu^{\alpha}$, where S$_{\nu}$ is the flux
density at a frequency $\nu$) during the flare in the harder state (consistent
with observations of radio flares in NSs: e.g. Cooke \& Ponman 1991; Penninx
et al. 1993; Hjellming et al. 1990a,b; Fomalont, Geldzahler \& Bradshaw 2001).
For the identifications of the timing components in the power spectra
(Fig.~1), we followed the detailed timing analysis in Schnerr et al. (2003).

{\em GX~339-4:} GX~339-4 is one of the two BHs to reveal a correlation between
radio and X-ray flux in hard state over a few orders of magnitude (Corbel et
al. 2001; the other is V404 Cyg: Gallo et al. 2003), and the only one to have
PCA/RXTE observations, simultaneous with ATCA radio observations, which can
trace smoothly the X-ray timing behaviour of the source in hard state over an
order of magnitude in radio luminosity. In order to compare the relations
between radio emission and X-ray timing among NSs and BHs, we have analysed
the simultaneous observations of the BH GX~339-4 reported in Corbel et
al. (2000; see Table~1), during the hard state, and which have an optically
thick radio spectrum (i.e. steady compact jet). The identification of the
features has been made using Psaltis et al. (1999), Belloni et al. (2000) and
Klein-Wolt et al. (2005), in prep. 

Note that previous radio/X-ray timing analysis of other BH XRBs shows in the
`peculiar' BH GRS~1925+105 a relation between the radio states and the QPO
centroid frequency, i.e. the frequency is lower during the so-called radio
plateau state where the source has a higher optically thick radio flux (Muno
et al. 2001), and no evidence for correlations in Cyg X-1, which is almost
persistently in the hard state, but steadily at high X-ray luminosties close
to the the soft state transition (Pottschmidt et al. 2003).

\section{Results}\label{results}

We find significant correlations between the radio luminosity L$_{R}$ (at
8.5~GHz and calculated using the distances listed in
Table~\ref{qpo-radio_tab2}) and the X-ray timing features L$_{h}$ and L$_{b}$
in NS XRBs (Fig.~2a,b,c) and L$_{\ell}$ in the BH GX~339-4
(Fig.~2a',b',c'). L$_{h}$ and L$_{b}$ in the NSs and L$_{\ell}$ and
L$_{break}$ in the BH have been chosen because they are the components that
are almost always present in the power spectra thus allowing a comparison
among observations and sources.

Spearman rank tests give positive correlations for L$_{R}-\nu_{h}$ (99.2\%)
and L$_{R}-\nu_{b}$ (99.1\%) without considering ms accreting X-ray pulsars
(i.e. with 4U~1728--34, 4U~1820-34, Ser X-1, MXB~1730--335, GX~13+1). Adding
also the ms accreting X-ray pulsars (IGR~J00291+5934 and SAX~J1808.4--3658,
not shown in Fig.~2 and discussed in more detail in \S~\ref{discussion:ms}) we
still found a significant correlation for L$_{R}-\nu_{b}$ (99.3\%), but only
marginal for L$_{R}-\nu_{h}$ (97.4\%). 

As expected from the negative correlations found between the characteristic
frequency and the rms of the Lorentzian components in the power spectra of NS
XRBs (e.g. van Straaten et al. 2002, 2003, 2005), we also found a negative
correlation between L$_{R}$ and the rms of L$_{h}$ (99.1\% with atoll sources
and 99.7\% with also ms X-ray pulsars) and the rms of L$_{b}$ (98.8\% with
atoll sources and 99.7\% with also ms X-ray pulsars).

We fitted all these correlations with power-laws. The fits with a power-law
model give $\chi^{2}/{\it d.o.f.}$ not better than $\sim 4$ (considering
errors on both variables), however, due to a lack of physical models to
describe such correlations, the derived fitting parameters, like the power-law
spectral index, are useful to quantify the relations of these observational
quantities. To quantify the likelihood of the power-law model to describe the
correlations, we quote also the value of the linear coefficient $r$ calculated
for each correlation in a logarithmic scale. In order to be more conservative,
we chose to perform the fit considering the atoll sources without the
`peculiar' atoll source GX~13+1. In this way we can trace the behaviour of the
other sources (GX~13+1, but also of the ms X-ray pulsars) with respect to them
(Spearman rank tests without GX~13+1 and the ms accreting X-ray pulsars gives:
for L$_{R}-\nu_{h}$ a significance of 98.7\%, for L$_{R}-\nu_{b}$ a
significance of 98.9\% and for L$_{R}-$rms of L$_{h}$ a significance of
97.7\%).  In the following we quote the fitting-parameters' values with
$1\sigma$ errors derived using only errors on L$_{R}$ (similar values are
derived using errors on both variables).

Fits with a power-law L$_{R}=A\times\nu_{h/b}^{\Gamma}$ of the radio
luminosity as a function of the timing frequencies give, for $\nu_{h}$ a slope
of $\Gamma=1.30\pm0.10$ and a normalization $A=10.8\pm3.8$, and for $\nu_{b}$
a $\Gamma=0.63\pm0.04$ and $A=197.1\pm17.0$. Note that the L$_{R}-\nu_{b}$
correlation looks more `bimodal' than L$_{R}-\nu_{h}$ (Fig.~2,c; $r=0.85$ and
$r=0.93$, respectively, for L$_{R}-\nu_{b}$ and L$_{R}-\nu_{h}$).

A fit with a power-law of L$_{R}$ with the rms of the timing components give,
for the rms of L$_{h}$: $\Gamma=-2.32\pm0.14$ and
$A=7.303(\pm2.255)\times10^{4}$ with a correlation coefficient $r=-0.87$; and
for the rms of L$_{b}$: $\Gamma=-1.66\pm0.20$ and $A=7.65\pm3.81$ with a
correlation coefficient $r=-0.66$.

Correlations similar to those found for $\nu_{h}$ and for the break frequency
in NSs hold also for the BH GX~339-4 in the hard state. In Fig.~2a' we show
the radio luminosity at 8.5~GHz as a function of the characteristic frequency
$\nu_{\ell}$ in the BH GX~339-4 (see \S~\ref{discussion:NSvsBH} for a
discussion). The fit with a power-law gives $A=(4.88\pm0.11)\times10^{3}$, and
a slope of $\Gamma=1.37\pm0.02$ (slope consistent with being the same as with
$\nu_{h}$ in atoll NSs). The rms of L$_{\ell}$, as in NSs, decreases with
radio luminosity and a power-law fit gives $A=3.54(\pm0.45)\times10^{7}$ and a
slope of $\Gamma=-2.38\pm0.03$ (with a correlation coefficient $r=-0.90$;
Fig.~2b'). This slope is also consistent with the $\Gamma$ found for the rms
of L$_{h}$ in atoll NSs.  A marginally significant correlation (95\%) has been
found between the radio luminosity and the break frequency $\nu_{break}$
(Fig.~2,c'). A power-law fit (without the upper limit point) gives
$\Gamma=0.87\pm0.02$ and $A=2.59(\pm0.99)\times10^{5}$, with a linear
correlation coefficient of $r=0.88$. 

As a caveat we note that, in NSs, in the case of a transient outburst, X-ray
and radio emission may have been decoupled. X-ray timing features, being
related to the disc properties, should be a more reliable tracer of the
compact steady continuous replenished jet, rather than of the evolution of a
transient optically thin jet, at least after the jet has been launched and
thus decoupled from the disc. In this context, this caveat specifically
applies to the ms accreting X-ray pulsar SAX~J1808.4--3658 detected in radio
only during X-ray outbursts that do not, in fact, follow the L$_{R}-\nu_{h}$
correlation found for the other sources (Fig.~\ref{ms}; see
\S~\ref{discussion:ms}). The fact that the data of the (likely) transient jet
in the Rapid Burster (and actually also the ms accreting X-ray pulsar
IGR~J00291+5934 which is in the last part of the X-ray outburst decay) instead
do lie on the radio-X-ray timing correlations may suggest a continuity in
power between the compact and the transient jet at launch (see discussion in
Fender, Belloni \& Gallo 2004 for BHs).

\section{Discussion}

We have analysed all the available PCA/RXTE observations coordinated with
radio, of atoll NSs and ms accreting X-ray pulsars with a detected radio
counterpart.

We find significant correlations between the radio luminosity at 8.5~GHz and
X-ray timing features. We compared these relations in NSs with those of the BH
GX~339-4 in the hard state, for which we could smoothly follow the radio
behaviour and the high-resolution X-ray timing variations over an order of
magnitude in radio luminosity. We found that similar relations hold in the two
different classes of sources. In particular we observe: in atoll NSs
correlations between radio luminosity and L$_{h}$ and between radio luminosity
and the break frequency component L$_{b}$, in BHs correlations between radio
luminosity and L$_{\ell}$ and (marginally) between radio luminosity and the
break frequency L$_{break}$. In the following we will discuss some
implications of these results and suggest possible scenarios, which 
require testing using a larger sample.

\subsection{Atoll-type NSs} 
\label{discussion:atolls}

The phenomenon of the parallel tracks in the X-ray luminosity vs. kHz QPO
frequency plot for NSs (e.g. M\'endez et al. 1999; see van der Klis 2005 for a
review) may be explained using a single time dependent physical quantity,
usually inferred as the mass accretion rate in the disc $\dot{M}_{d}$, if we
consider the kHz QPO frequency to be governed by the balance between this
quantity and luminosity L$_{X}$, while the luminosity responds to both this
quantity and its time-averaged variations (van der Klis 2001). The QPOs seem
to be a good tracer of the disc geometry as determined by the balance between
$\dot{M}_{d}$ and L$_{X}$, in the regions very close to the compact object
where also jets are thought to originate (see also below).  van der Klis
(2001) also suggested that one possible mechanism for balancing the energy
budget is by removing material from the binary in the form of a
mildly-relativistic jet, so that the accretion rate on to the compact object
is no longer the same as the mass transfer rate through the disk.

All the variability components in the power spectra follow a universal scheme,
when plotted against the upper-kHz QPO. Therefore the disc geometry may be
inferred also by low-frequency timing features. In particular in atoll
sources, a tight correlation exists between the centroid frequency of the
upper-kHz QPO and that of (the radio power tracer) L$_{h}$: the best-fit
power-law is $\nu_{h,0}\propto\nu_{u,0}^{2}$ (van Straaten, van der Klis \&
M\'endez 2003), or $\nu_{h}\propto\nu_{u}^{2.4}$ using $\nu_{max}$ (van
Straaten, van der Klis \& Wijnands 2005). If we assume that $\nu_{u}$ is
related to an orbit in the disc at an inner radius R$_{in}$, the fact that the
radio jet power increases with $\nu_{h}$, may be explained by a scenario in
which the jet particles and magnetic field lines may be `squeezed' as the disc
moves inwards. (The jet power possibly increases until the magnetic field
lines of the jet interact with the magnetic field of the NS surface; see also
below in \S~\ref{discussion:ms}).

\subsection{Millisec accreting X-ray pulsars}
\label{discussion:ms}

\begin{figure}
\begin{tabular}{c}
\psfig{figure=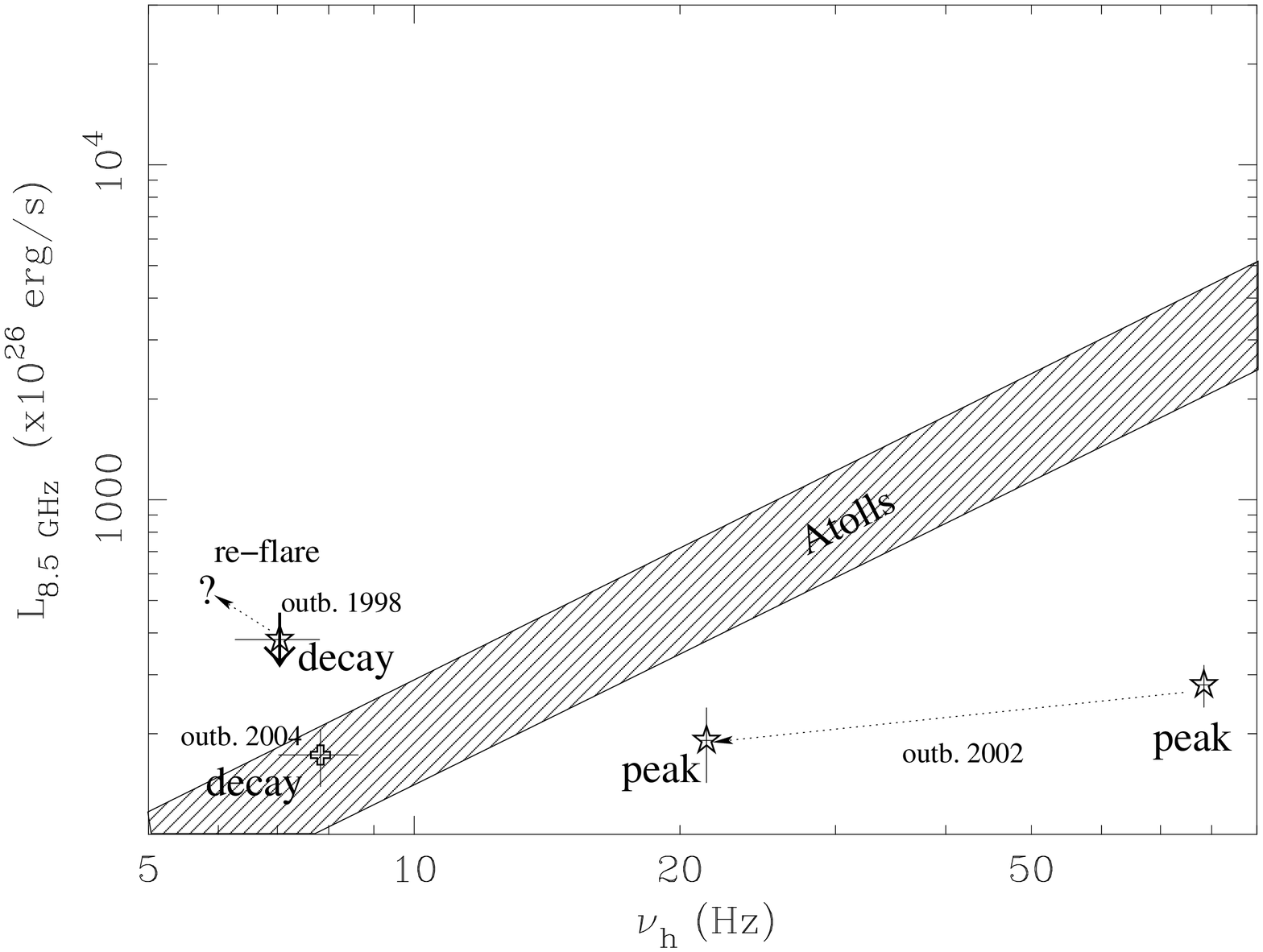,width=8cm,angle=0}\\
\psfig{figure=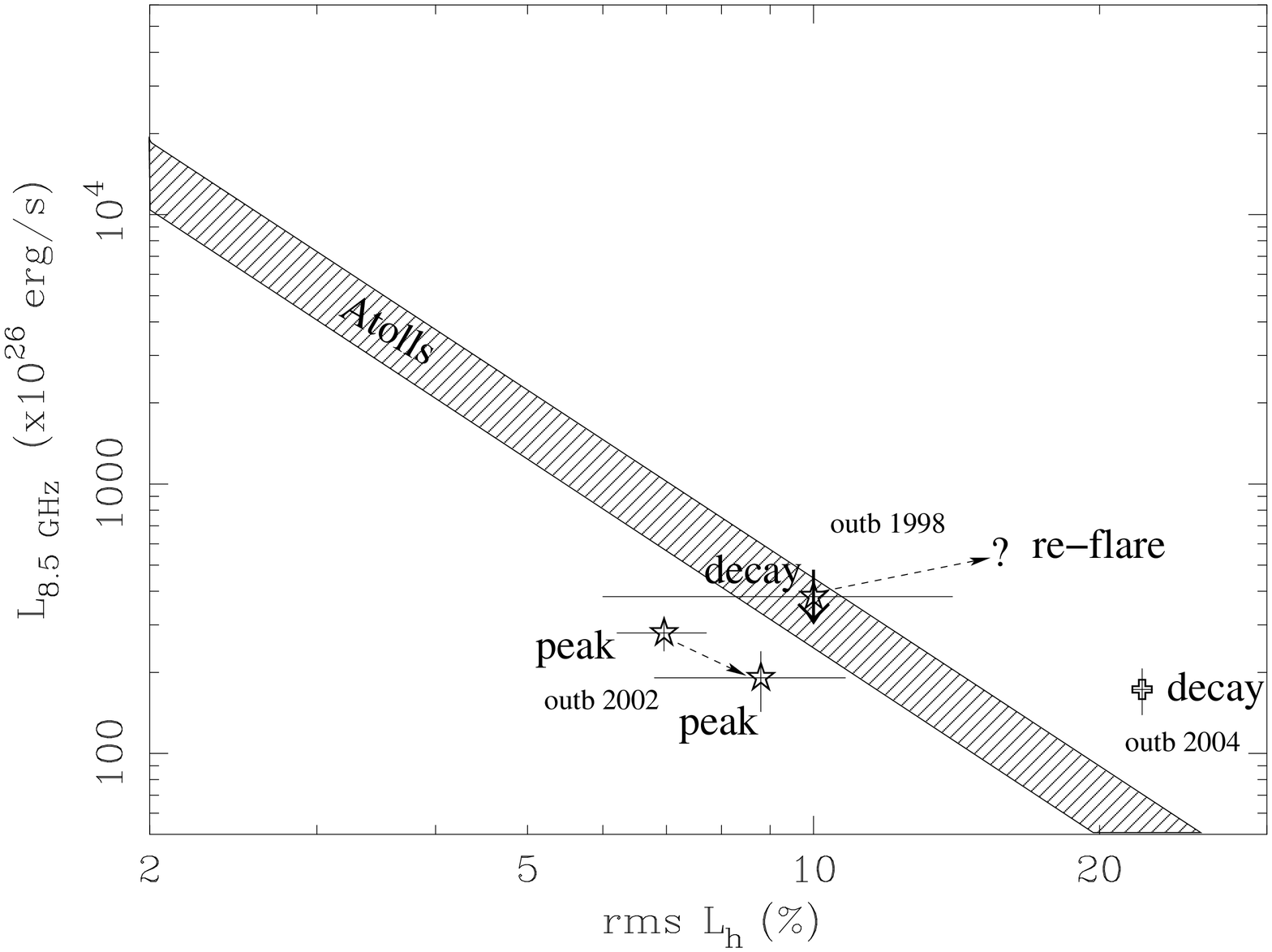,width=8cm,angle=0}\\
\psfig{figure=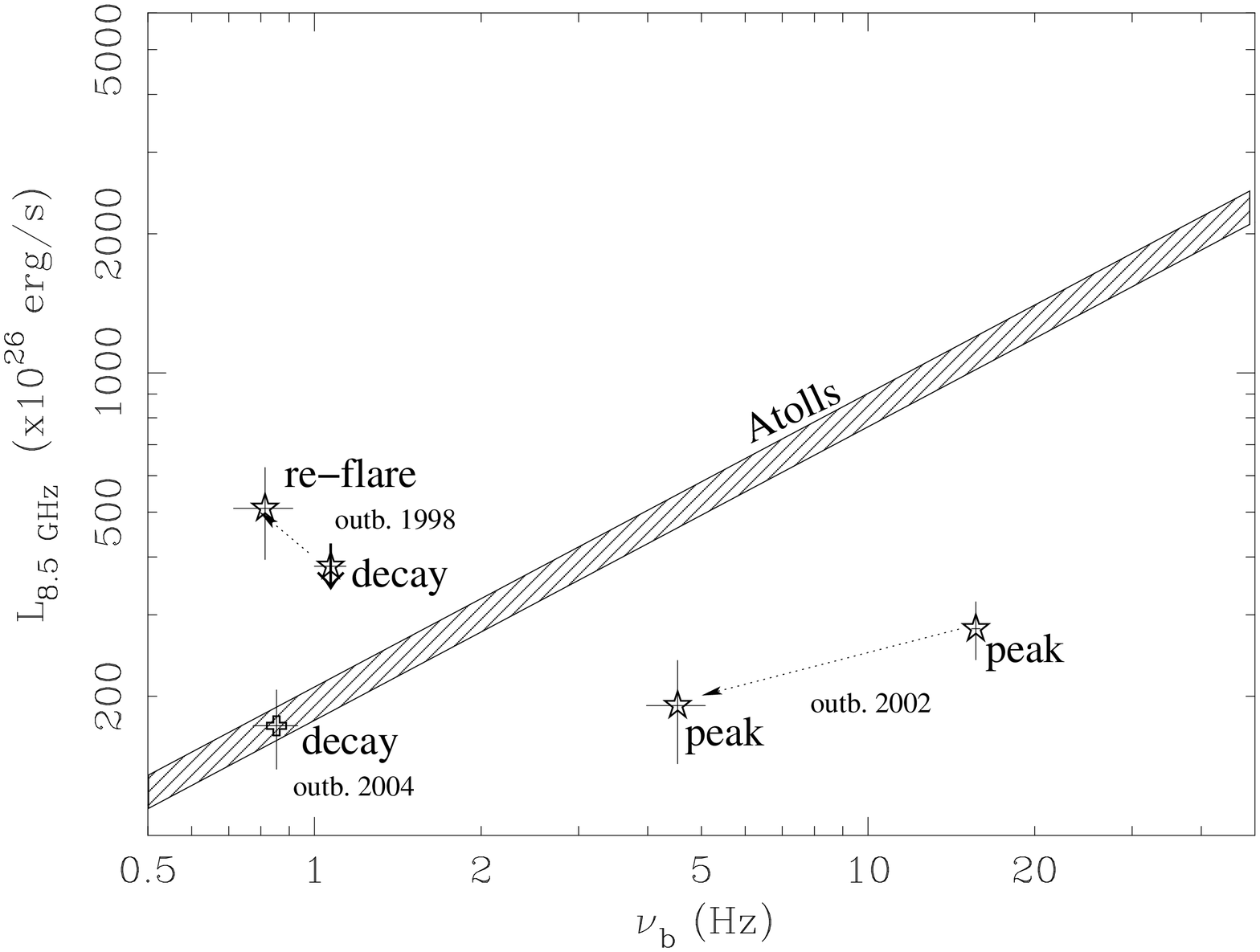,width=8cm,angle=0}\\
\end{tabular}
\caption{
Radio luminosity at 8.5~GHz as a function of $\nu_{h}$ (upper panel), the rms
of L$_{h}$ (middle panel) and $\nu_{b}$ (lower panel) for the ms accreting
X-ray pulsars IGR~J00291+5934 (open cross) and SAX~J1808.4--3658 (stars). The
gray region represent the power-law fit to the atoll sources considering
$1\sigma$ errors on the normalization A only
(e.g. S$_{8.5}$=$A\times(\nu)^{\Gamma}$; see \S~\ref{results}). The question
mark indicates an uncertainty on $\nu_{h}$, because since no L$_{h}$ could 
be identified on 1998 April 27, we used the value of the X-ray observation
closest to that day and for which $\nu_{b}$ was consistent with the one on
1998 April 27 (i.e. group number 21 in van Straaten et al. 2005; see also
\S~\ref{sample}). The arrows indicate the chronological sequence of the
observations during each X-ray outburst.}\label{ms}
\end{figure}

The observations of the ms accreting X-ray pulsar IGR~J00291+5934, taken in
the last part of the decay of the X-ray outburst on 2004, also lies on the
radio luminosity/X-ray timing frequency correlations of atoll sources (adding
this source the L$_{R}-\nu_{b}$ and L$_{R}-\nu_{h}$ rank correlations are
significant to the 99.6\% and 99.7\% level, respectively). On contrary, the
observations of SAX~J1808.4--3658 seem to diverge from the radio/X-ray
coupling found in atoll sources and IGR~J00291+5934, especially from the
L$_{R}-\nu_{h}$ relation (see \S~\ref{results}; note that the ms X-ray pulsars
are still consistent with the rank correlations between radio luminosity and
the rms of the X-ray timing features of atolls sources; see also Fig.~3,
middle panel). The radio upper limit of SAX~J1808.4--3658 has been derived on
1998 April 26 from an observation during the last part of the decay of an
X-ray outburst, as for IGR~J00291+5934, and this value is also consistent with
the correlations of the atolls. The observations that do not lie on the radio
luminosity/X-ray timing frequency correlation are those of SAX~J1808.4--3658
taken during the peak of the outburst in 2002 and the one that apparently
re-flared on 1998 April 27. Although the lack of observations does not allow a
complete understanding yet of the behaviour of the simultaneous radio/X-ray
emission of ms X-ray pulsars throughout the evolution of an X-ray outburst,
based on the present data we can start to make some remarks and speculations.

Focusing on the radio luminosity/X-ray timing frequency relations
(Fig.~\ref{ms}, upper and lower panels) we see that when the frequency is high
($\nu_{h}>20$; $\nu_{b}>4$) the radio luminosity is below the one expected
from atoll sources, while it is above during the re-flaring when the source
show lower timing frequencies. Furthermore, we see that in the 2002 outburst,
in two days the timing frequencies change of a factor of four while only a
little decrease is observed in radio luminosity, i.e. an increase in radio
loudness. This may be an indication of an actual decoupling of disc and jet
during the outburst, thus signature that a discrete jet has been launched. The
fact that when the timing frequencies are higher the source is less radio loud
than at lower timing frequencies, may be due to the relative importance of the
magnetic field of the NS when the disc is getting closer to the NS during the
outburst. For the two observations in 2002 we have a direct measurement of the
upper kHz QPOs: L$_{u}\sim700$~Hz on October 16 and L$_{u}\sim400$~Hz on
October 18 (see also van Straaten et al. 2005). If we relate this frequency to
the Keplerian orbit in the disc at a radius R$_{in}$, for a 1.4~M$_{\odot}$ NS
we derive R$_{in}\sim70$~km in the October 16 observation and
R$_{in}\sim100$~km in the October 18 observation. For the observation of 1998
April 27, using the relation $\nu_{h}\propto\nu_{u}^{2.4}$ we derive
R$_{in}\sim140$~km. In ms accreting X-ray pulsars the magnetic field of the NS
might be higher than in atoll NSs (e.g. Chakrabarty 2005), and for
SAX~J1808.4--3658 it has been estimated to be $\sim 10^{8}-10^{9}$~G (Psaltis
\& Chakrabarty 1999). The (yet poor) observations may suggest that below, say,
$\sim10^{2}$~km from the NS, the magnetic field, interacting more strongly
with the regions of the disc where the jet is formed - closer to the compact
object as the jet gets more powerful as suggested by the correlation with the
X-ray timing frequencies - suppresses in some way the power of the jet
(possibly by interfering with a magnetic expulsion mechanism; see also
\S~\ref{discussion:atolls}).
We can have a rough estimate of the radius at which this happens if we assume
that this interaction is strong when this region is close to the Alfv\'en
radius R$_{A}$.  Assuming for SAX~J1808.4--3658 a magnetic field at the
surface of $\sim5\times10^{8}$~G (Psaltis \& Chakrabarty 1999), a luminosity
during the decay of the 1998 outburst on April 27 of
$\sim3\times10^{35}$~erg/s [from Gilvanov et al. (1998), using a distance of
2.5 kpc (in 't Zand et al. 2001)], and a radius of the NS of $\sim10$~km
(Burderi \& King 1998), we obtain R$_{A}\sim110$~km, and slightly smaller
radii for the other observations (see e.g. Frank, King \& Raine 2002 and
references therein). This radius is consistent with the inner radius we have
estimated with kHz QPOs' frequencies, therefore suggesting that the magnetic
field of the NS, through its interaction with the innermost regions of the
accretion disc, might play a role also in (e.g. inhibiting) the production of
the jet.

\subsection{NSs and BHs}
\label{discussion:NSvsBH}

Casella, Belloni \& Stella (2005) found a clear association between a
low-frequency QPO, the so-called type-C QPO in BHs (e.g. Wijnands et al. 1999;
Remillard et al. 2002; Casella et al. 2004) and the horizontal-branch
oscillation (HBO) in Z-type NSs, suggesting a similar physical origin for
these two timing features. van Straaten et al. (2003) already compared the
low-frequency features of atolls with those of Z sources, and, based upon the
correlation with the frequency of the kHz QPOs, identified the HBO in Z
sources with L$_{h}$ in atoll sources. [Note that, a correlation between radio
power and the position in the CD has been found in Z sources, and it is
precisely in the HB that Z sources are more radio loud (Penninx et al. 1988;
Hjellming et al 1990a,b).] In BHs, Psaltis et al. (1999; see also Fig.~12 in
Belloni et al. 2002) found a tight almost {\em linear} correlation between the
frequencies of the narrow low-frequency QPO L$_{LF}$ (i.e. also the type-C
QPO) and a broader component that they called L$_{\ell}$ (reminiscent of the
lower kHz component in NSs).
Using these known relations we can (linearly) associate L$_{h}$ of NSs with
L$_{\ell}$ of BHs (where the choice for these two features have been dictated
by availability in the power spectra of our sample; see \S~\ref{results}). The
fact that we find for these two features the same scaling with radio
luminosity supports the idea that the same physics lie behind their origin.

Since all the compact jet models predict that the total jet power L$_{J}$ and
the power radiated in radio L$_{R}$ are related as
L$_{R}\propto$L$_{J}^{1.4}$ (e.g. Blandford \& Konigl 1979; Falcke \& Biermann
1996; Falcke, Markoff \& Fender 2001), the finding that the radio luminosity
scales with $\nu_{h}$ and $\nu_{l}$, in NSs and BHs respectively, with a slope
of $\Gamma\sim1.3-1.4$, would imply an almost linear correlation of L$_{J}$
with these timing features.

It is interesting to note that if we consider the frequencies of the timing
components correlated with the radio luminosity (i.e. $\nu_{h}$ for NSs and
$\nu_{\ell}$ for the BH GX~339-4) as originating from Keplerian motion of
matter in the disc, from the dynamical timescales observed we derive a
distance to the compact object of $\sim100-300$~R$_{Schw}$ and
100-500~R$_{Schw}$ (where R$_{Schw}=2GM/c^2$ is the Schwarzschild radius),
respectively in NSs and in GX~339-4. These radii are consistent with those
derived from high-resolution radio imaging of the active galaxy M87 (Junor,
Biretta \& Livio 1999) and consistent with the idea that jets are formed very
close to the compact object.
 
The relations between the radio luminosity and the break frequency are even
more intriguing, because such features have been observed also in AGN
(e.g. McHardy et al. 2004). Therefore, these correlations, if further
confirmed especially with a larger sample of BHs, open the possibility of the
existence of a new `fundamental plane' for XRBs and AGN (Merloni, Heinz
\& Di Matteo 2003), where the dimensions are the mass of the compact object,
the radio luminosity and the characteristic frequency of a timing feature
(e.g. break), the last being, contrary to the X-ray luminosity (see Merloni et
al. 2003 and Falcke, K\"ording, Markoff 2004), independent from the distance
to the source. X-ray timing features can be the key features to finally find
the common physical link between the accretion disc and the jet radio emission
in NSs and BHs of all masses.

\section*{Acknowledgements}
We thank the anonymous referee for her/his useful suggestions. We acknowledge
useful conversations with Phil Uttley. SM would like to thank Rudy Wijnands,
Tomaso Belloni, Marc Klein-Wolt, Diego Altamirano and Manuel Linares for very
helpful discussions.

\end{document}